\documentclass[runningheads]{llncs}
\usepackage{graphicx}
\usepackage{svg}
\usepackage{listings}
\usepackage{verbatim}

\begin{document}
\title{Duplication Detection in Knowledge Graphs: Literature and Tools}

\author{Elwin Huaman \and Elias K\"arle \and Dieter Fensel}
\authorrunning{E. Huaman et al.}
\institute{Semantic Technology Institute, Innsbruck, Austria\\
\email{\{firstname.lastname\}@sti2.at}}
\maketitle              
\begin{abstract}
In recent years, an increasing amount of knowledge graphs (KGs) have been created as a means to store cross-domain knowledge and billion of facts, which are the basis of costumers' applications like search engines. However, KGs inevitably have inconsistencies such as duplicates that might generate conflicting property values. Duplication detection (DD) aims to identify duplicated entities and resolve their conflicting property values effectively and efficiently.
In this paper, we perform a literature review on DD methods and tools, and an evaluation of them. Our main contributions are a performance evaluation of DD tools in KGs, improvement suggestions, and a DD workflow to support future development of DD tools, which are based on desirable features detected through this study.

\end{abstract}
\section{Introduction}
\label{sec:introduction}

\begin{figure}
    \includegraphics[width=\textwidth]{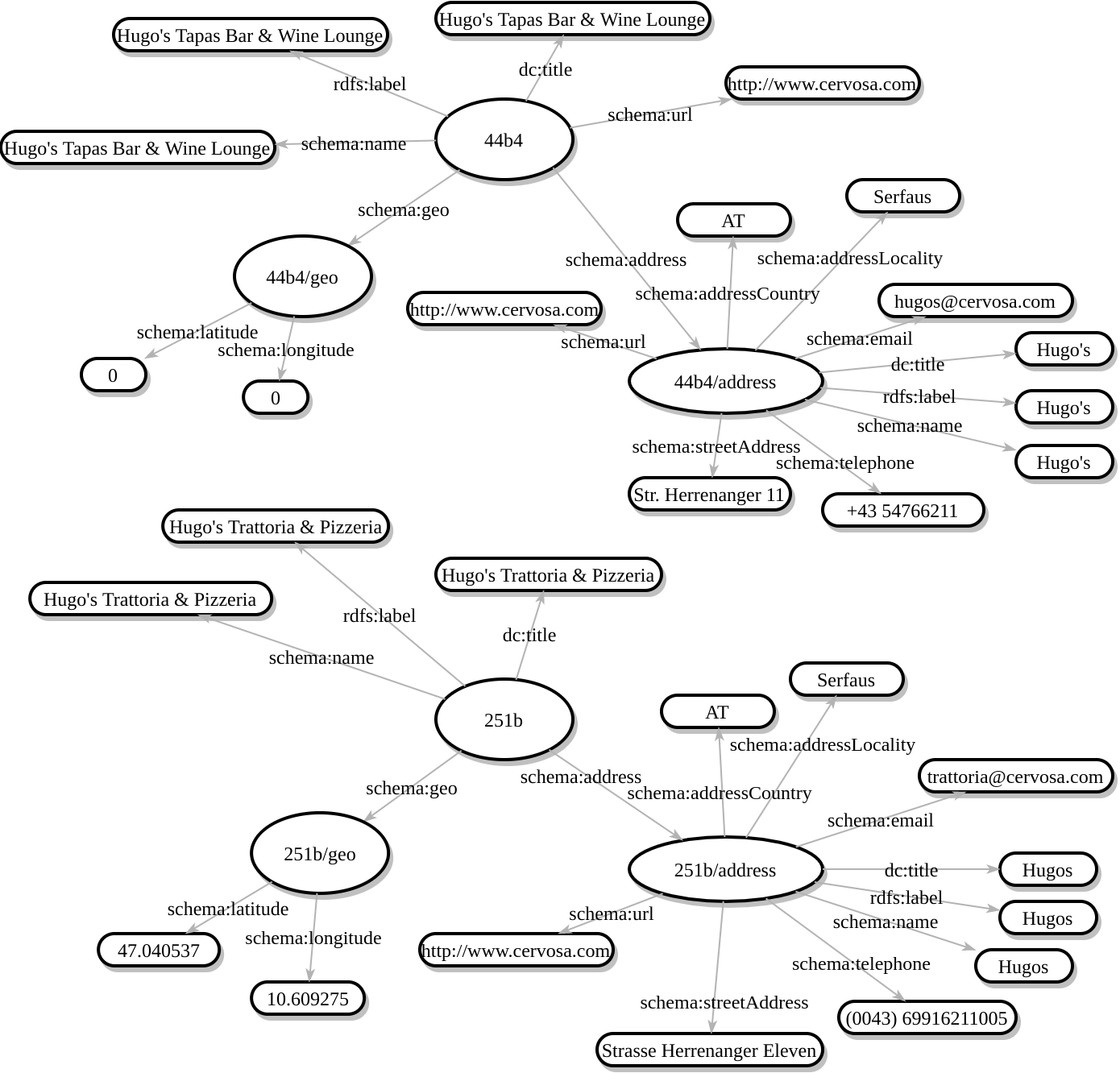}
    \caption{Two instances that are contained in a KG. They show potential duplication sources that can occur between them.} 
    \label{fig:duplicates}
\end{figure}

Over the last years, the number and variety of created knowledge graphs (KGs) have increased~\cite{Hogan2020}. KGs aim to store cross-domain knowledge and billion of facts (e.g. Amazon Product Knowledge Graph~\cite{Dong2019}, Google’s Knowledge Graph, Bing Knowledge Graph, eBay's Product Knowledge Graph~\cite{Noy2019}), and to provide structured data for costumers' applications such as search engines and question answering systems. However, KGs generated and used by information systems inevitably have inconsistencies, such as duplicates that can lead to generating wrong instance-, property value-, and  equality assertions~\cite{FenselSAHKPTUW2020}. The presence of such duplicates (See Figure~\ref{fig:duplicates}) may compromise the outcome of business intelligence applications. Hence, it is crucial and necessary to explore efficient and effective semi-automatic methods and tools for tackling the duplication detection (DD) in KGs. In other words, evaluating existing DD methods and tools.

In this paper, we present a literature review of DD tools aiming to find effective and efficient ones. We state the DD problem by describing its origin and main issues (or tasks). Afterwards, we discuss the DD literature, showing methods and tools. Then, we evaluate the DD tools and discuss their results. Furthermore, we provide improvement suggestions and a DD workflow.

In order to have a comprehensive review of the current DD literature and tools, first, we state the DD problem identifying two main issues (or tasks): 1) identifying and resolving duplicates, and 2) resolving conflicting property value assertions. Afterwards, we identify state-of-the-art tools for each task. Then, we define our evaluation approach that consists on six research questions, selecting a dataset to be evaluated, generating a benchmark (or golden standard), selecting tools based on three criteria, and executing the selected tools. Finally, we discuss their performance by answering the defined research questions.

The remainder of this paper is organized as follows, Section 2 states the DD problem including its origin and main issues. Section 3 defines the evaluation approach. Then, the evaluation of DD tools regarding their functionality and performance is described in Section 4. Evaluation results are given in Section 5. Finally, Section 6 summarises the conclusions and future work.

\section{Problem Statement}
\label{sec:problem-statement}
Duplication detection (DD) in knowledge graphs (KGs) context is a problem of highly practical relevance to ensure the quality of KGs, and there is a need for effective\footnote{Achieving the comparison of all property values of entities.} and efficient\footnote{Optimizing the speed and used resources to compare large number of entities.} frameworks to tackle the problem. Before going into the details of such frameworks in Section \ref{sec:evaluation}, we first state the duplication origin, afterwards we describe the duplicate detection issue.

\subsection{Duplication Origin}
\label{subsec:duplication-origin}
The main reasons of having duplicates in a knowledge graph (KG) lie on: a) the entered of the same entity multiple times into the KG, and b) during the integration into the KG of multiple data sources which have different representation of the same entity. The first case might happen when a KG is crowd-source-based and a user does not verify if the entity she is entering is already in the KG (e.g. a user on Wikidata can enter two times the same entity\footnote{See more: \url{https://www.wikidata.org/wiki/Wikidata:True\_duplicates}}). The second case happens when integrating different knowledge sources which have heterogeneous schemata (see Figure \ref{fig:duplicates}).

The error sources are very diverse, we distinguish them in two categories a) property-value, where there exist: typos like \textit{Hugos} versus \textit{Hugo's}; different data entry like \textit{Str. Herrenanger 11} versus\textit{ Strasse Herrenanger Eleven} versus \textit{11 Str. Herrenanger} or \textit{Serfaus, AT} versus \textit{Serfaus}; missing values like geo property values \textit{0,0} versus \textit{47.040537,10.609275}, and b) heterogeneous property names where the schema used to represent an entity varies, for instance \textit{schema:address} versus \textit{schema:streetAddress}) or \textit{rdfs:label} versus \textit{schema:name} (see Figure \ref{fig:duplicates}).

\subsection{Duplication Detection (DD)}
\label{subsec:duplication-detection}
DD involves two main issues: a) the effective comparison of property-values of entities, and b) the efficient comparison of large number of entities, which make the DD task hard, furthermore, we should add the complexity of dynamic data, since an entity can be represented differently over a period of time. For instance, the telephone number of a restaurant can change. Besides that, given the size nature of KGs the detection of duplicates becomes even more complex. Therefore, two main problems arise when dealing with DD in KGs context: achieving both a high effectiveness and a high efficiency of the DD process. In other words: how do we find duplicated entities in a KG and how do we resolve the conflicting property values between duplicates in an effective and efficient way?

For producing a more consistent, accurate, and useful KG, two issues must specially be tackled, these issues (or tasks) have been defined based on a maximal simple knowledge representation formalism in \cite{FenselSAHKPTUW2020} as follows:

\begin{enumerate}
    \item \textbf{Identifying and resolving duplicates}: Deriving new $isSameAs(i\textsubscript{1}, i\textsubscript{2})$ assertions and aligning the descriptions of these two identifiers, which are symetric, reflexive, transitive, and semantically equivalent. Also, every statement remains true or false when replacing $i\textsubscript{1}$ by $i\textsubscript{2}$ (and vice versa).
    \item \textbf{Resolving conflicting property value assertions}: Handles situations such as $P(i\textsubscript{1}, i\textsubscript{2})$, and $P(i\textsubscript{1}, i\textsubscript{3})$, and $i\textsubscript{2} =/= i\textsubscript{3}$, and $P$ has a unique value constraint. Furthermore, situations such as domain/range violations and having multiple values for a unique property (this refers to error detection and correction).
\end{enumerate}

DD tools try to identify all pair instances $(i\textsubscript{1},i\textsubscript{2})$ such as $i\textsubscript{1}$ and $i\textsubscript{2}$ are duplicates. The result of the DD tools is represented as a set of $isSameAs(i\textsubscript{1}, i\textsubscript{2})$ assertions. Furthermore, the DD tools typically computes a similarity score $sim \in [0,1]$ to the results. Where $0$ is the minimum degree of similarity regarding a metric and a value $1$ is the maximum degree.

DD performance is characterized by its effectiveness on comparing property-values of instances and its efficiency on comparing large number of instances. Therefore, achieving a high performance is challenging due to the large size of KGs, which typically cover multiple domains of billions of facts. Previous studies compare different similarity metrics \cite{BilenkoM2003} and tools \cite{BatiniS2006} and survey different comparison approaches \cite{LalithaMK2012} in order to identify their usefulness. Author-name disambiguation, Coreference Resolution, Deduplication, Entity Linking, Identity Resolution, Object Consolidation, Record Linkage, and Schema Matching are different conceptual frames of the same problem \cite{GetoorM2012}. The flexibility of describing data in KGs is of course an advantage to store heterogeneous data sources, however this heterogeneity causes still error sources (see Section \ref{subsec:duplication-origin}).

Our goal is to provide a literature review on DD that can encourage the reuse, improvement or development of approaches, methods, and tools.

\section{Literature Review}
\label{sec:literature}
In this section, methods and tools will be discussed. First, we describe methods for identifying duplicates, afterwards, we summarise existing tools for detecting and resolving duplicates, and tools for resolving conflicting property values.

\subsection{Methods}
\label{subsec:methods}
The current literature regarding DD frameworks have been studied and some approaches and methods have been proposed. For instance, approaches that address the DD in specific domains \cite{GiannopoulosSMKA2014} \cite{RaimondSS2008} and methods that apply association rule mining \cite{HippGN2000}, crowd-sourced data \cite{GetoorM2013}, graph-oriented \cite{KorulaL2014}, property-based \cite{HoganHD2007}, network metrics~\cite{IdrissouHB2018}, Support Vector Machine \cite{HansenBFGGKRV2015}, string similarity measures \cite{BilenkoM2003}, and topic modeling~\cite{JuAJHYM2016} techniques. More methods can be found in~\cite{ShaoxiongSEPP2020}.

\subsection{Tools}
\label{subsec:tools}

DD is not a straightforward task and several tools have been developed. The tools mentioned below are very heterogeneous in the sense of they have different setups, accept only a few file formats, they were programmed using different programming languages, they address DD in one dataset or between two datasets (i.e. interlinking of data sources), and more.

\textbf{Identifying and resolving duplicates} is identifying where two or more records in a single or various KGs are referring to the same entity and linking those. The tools found during the review of the literature are ADEL \cite{PluTR2017}, Dedupe \cite{BilenkoM2003}, DuDe \cite{Draisbach2010}, Duke \cite{GarsholB2013}, Legato \cite{AchichiBT2017}, LIMES \cite{NgomoA2011}, SERIMI \cite{AraujoHSV2011}, and Silk \cite{VolzBGK2009}.

\textbf{Resolving conflicting property value assertions} or data fusion refers to handle for example situations such as the pair of duplicated entities have a different value for the same property, the state-of-the-art tools for tackling this task are FAGI \cite{GiannopoulosSMKA2014}, Sieve \cite{MendesMB2012}, and SLIPO Toolkit \cite{AthanasiouAGKKM2019}.

The most of the tools mentioned above need a previous configuration to start working, such as Silk and Sieve. Also, most of the approaches focus on an individual type of use cases (e.g. FAGI focuses on geospatial data). We also notice that none of these tools have handled both of the defined tasks and they are mostly focused on the detection of duplicates rather than on the resolution of the conflicting property values.

\section{Evaluation Approach}
\label{sec:evaluation}
In this section, we describe our evaluation approach, which starts by choosing a dataset to be used as input for the tools, afterwards we generate a benchmark based on the dataset for performance evaluation of tools, later we define a set of similarity metrics for the DD task, then we select a set of tools based on criteria, finally we setup and execute the tools. In order to perform this study, the following research questions (RQ) were conceived concerning DD tools:

\begin{enumerate}
    \item Are the identified tools able to detect duplicates in knowledge graphs?
    \item Do the identified tools provide cleaning and/or preprocessing techniques?
    \item Do the identified tools provide enough comparators?
    \item How the identified tools tackle conflicting property values?
    \item How flexible is the configuration of the identified tools?
    \item How scalable are the identified tools regarding large knowledge graphs?
\end{enumerate}

\subsection{Dataset Selection}
\label{subsec:dataset-selection}

In this section, we describe the dataset that will be used throughout this paper. The data used in this study is obtained from the MindLab Knowledge Graph (MKG)\footnote{\url{https://graphdb.sti2.at/repositories/mindlab}}. The MKG has been constructed by integrating data from heterogeneous sources, e.g. Mayrhofen, Seefeld, and Serfaus-Fiss-Ladis. Currently, the MKG has around 2,632,607 statements and 732 distinct classes.

\begin{minipage}{0.95\textwidth}
\begin{lstlisting}[basicstyle=\ttfamily\scriptsize, 
    caption={An excerpt of the restaurant dataset in turtle syntax that shows the properties used at the first hierarchical level to describe restaurant instances.}, 
    label={lst:seefeld-restaurant}, frame=tlrb]
@prefix seefeld: <https://graph.mindlab.ai/tvb-seefeld/feratel/> .
@prefix dc: <http://purl.org/dc/terms/> .
@prefix rdfs: <http://www.w3.org/2000/01/rdf-schema#> .
@prefix rdf: <http://www.w3.org/1999/02/22-rdf-syntax-ns#> .
@prefix schema: <http://schema.org/> .
@prefix xsd: <http://www.w3.org/2001/XMLSchema#> .

seefeld:a2ad4f06-a41d-4cbb-9a42-cdcee95d1d23 a schema:Restaurant ;	
  dc:title "Hotel Seespitz, Restaurant"@de , "Hotel Seespitz"@en ;
  rdfs:label "Hotel Seespitz, Restaurant"@de , "Hotel Seespitz"@en ;
  schema:name "Hotel Seespitz, Restaurant"@de , "Hotel Seespitz"@en ;
  schema:url "http://www.seespitz.at"^^xsd:anyURI ;
  schema:address seefeld:852e19a7-b71f-485a-a937-ea9a42cf5795/address ;
  schema:geo seefeld:a2ad4f06-a41d-4cbb-9a42-cdcee95d1d23/geo ;
  schema:image seefeld:02b8a90e-889e-4a3b-890d-12ab5f43c3fc .	
\end{lstlisting}
\end{minipage}

We fetch a dataset of all restaurant instances from the MKG to try the tools, the dataset contains 495 restaurant instances. In the following, we describe the properties used to describe restaurant instances in the dataset:

\begin{itemize}
    \item The restaurant instances are described using for example, the \textit{dc:title}, \textit{rdfs:label}, \textit{schema:address}, \textit{schema:geo}, \textit{schema:image}, and \textit{schema:name} properties.
    \item The values of \textit{schema:address}, \textit{schema:image}, and \textit{schema:geo} properties are instances of types, for example, the value of \textit{schema:address} is an instance of \textit{schema:PostalAddress} type.
    \item Mayrhofen, Seefeld, and Serfaus-Fiss-Ladis describe the address property (i.e. \textit{schema:address}) using more/less properties than each other, e.g. Mayrhofen uses eight properties and Seefeld uses twelve properties.
    \item If we have a close look on Listing \ref{lst:seefeld-restaurant} and Listing \ref{lst:seefeld-address-restaurant} - they are both part of the same restaurant instance (i.e. \textit{seefeld:a2ad4f06-a41d-4cbb-9a42-cdcee95d1d23 schema:address seefeld:852e19a7-b71f-485a-a937-ea9a42cf5795/address}) - we will find: conflicting property values in case of property \textit{schema:name} (e.g. "\textit{Hotel Seespitz, Restaurant}" versus "Hotel Seespitz****Superior"); redundant information (e.g. property \textit{schema:url}); and different properties representing the same value (e.g. \textit{purl:title}, \textit{rdfs:label}, and \textit{schema:name}).
\end{itemize}

Since not all of the selected tools accept the RDF format\footnote{Graph-based Data Model: \url{https://www.w3.org/TR/rdf11-concepts/}}, we have to convert the restaurant dataset into the CSV format. In order to use the same dataset with all tools, we reduced the number of properties used to describe restaurant instances to five commonly used properties\footnote{by Mayrhofen, Seefeld, and Serfaus-Fiss-Ladis to describe restaurants.}, which are: \textit{schema:name}, \textit{schema:url}, \textit{schema:streetAddress}, \textit{schema:latitude}, and \textit{schema:longitude}. Therefore, we define two queries that have been applied to the SPARQL endpoint of the MindLab repository\footnote{\url{https://graphdb.sti2.at/repositories/mindlab}}. The queries retrieve data that can be serialized as CSV and RDF format.

\begin{minipage}{1.0\textwidth}
\begin{lstlisting}[
    basicstyle=\ttfamily\scriptsize, 
    caption={An excerpt of the restaurant dataset in turtle syntax that shows a \textit{schema:PostalAddress} instance described by Seefeld.}, 
    label={lst:seefeld-address-restaurant}, frame=tlrb]
@prefix seefeld: <https://graph.mindlab.ai/tvb-seefeld/feratel/> .
@prefix dc: <http://purl.org/dc/terms/> .
@prefix rdfs: <http://www.w3.org/2000/01/rdf-schema#> .
@prefix rdf: <http://www.w3.org/1999/02/22-rdf-syntax-ns#> .
@prefix schema: <http://schema.org/> .
@prefix xsd: <http://www.w3.org/2001/XMLSchema#> .

seefeld:852e19a7-b71f-485a-a937-ea9a42cf5795/address a schema:PostalAddress ;
  dc:title "Hotel Seespitz****Superior" ;
  rdfs:label "Hotel Seespitz****Superior" ;
  schema:name "Hotel Seespitz****Superior" ;
  schema:url "http://www.seespitz.at"^^xsd:anyURI ;
  schema:addressCountry "AT" ;
  schema:addressLocality "Seefeld" ;
  schema:postalCode "6100" ;
  schema:streetAddress "Innsbruckerstr. 1" ;
  schema:email "info@seespitz.at" ;
  schema:faxNumber "(0043) 5212 2218 50" ;
  schema:telephone "(0043) 5212 2217" .
\end{lstlisting}
\end{minipage}

\subsection{Benchmark Generation}
\label{subsec:benchmark-generation}
In this section, we describe a benchmark (or gold standard) that will be used to evaluate the performance of DD tools. Benchmark datasets are very important for the evaluation of any existing or newly proposed tool, and for demonstrating tools applicability to a task. 

There are different ways to generate a benchmark, for instance, a benchmark can be generated using the KG itself and/or external KGs. There are methodologies that rely on the use of a gold standard dataset, a silver standard\footnote{“we assume that the given knowledge graph is already of reasonable quality” \cite{Paulheim2017}.} dataset, and on retrospective evaluation \cite{Paulheim2017} as a benchmark. Furthermore, the evaluation of the performance of DD tools can be measured in F-measure, Recall, and Precision. These metrics allow measuring the achieved performance of the tools, e.g., precision is the fraction of relevant instances among the retrieved instances.

\begin{figure}
    \centering
    \includegraphics[width=1\linewidth]{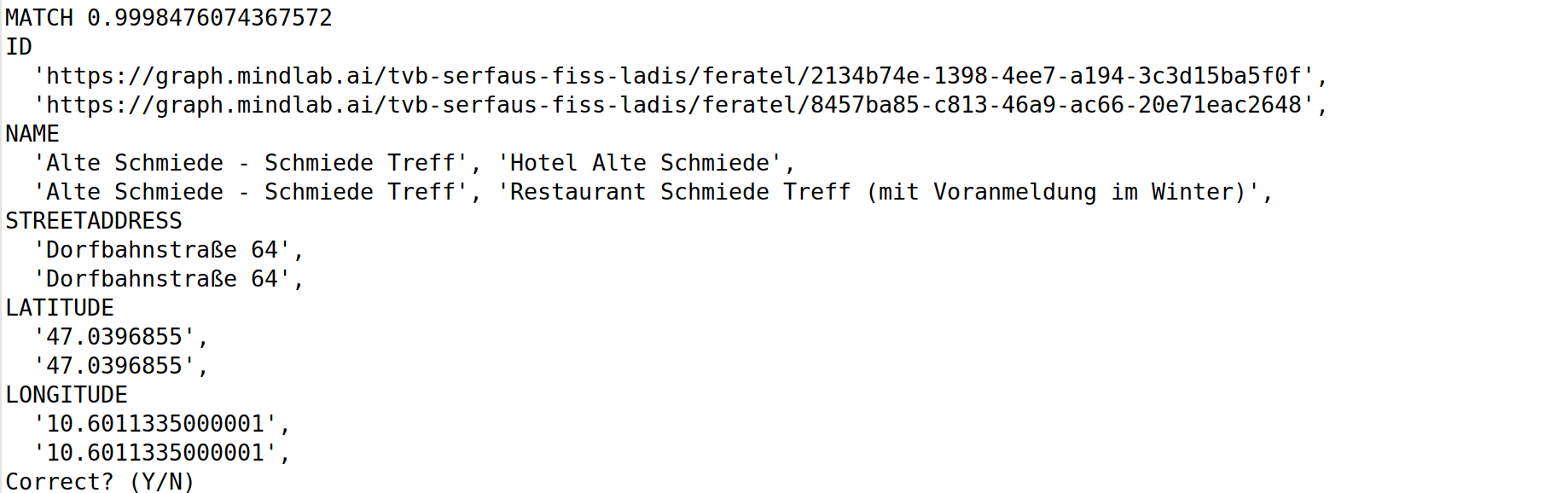}
    \caption{Labelling data as duplicates or non-duplicates. i.e. generating a gold standard.}
    \label{fig:gold-standard-generation}
\end{figure}

In this paper, we use the KG itself to generate a gold standard as a benchmark, which is a subset of the KG defined in Section \ref{subsec:dataset-selection}. The gold standard was labeled manually as duplicates or non-duplicates. We follow two steps: 1) given a similarity threshold of \textit{0.9} Duke shows potential matches through a command line interface (CLI) and a user should answer typing yes (Y) or no (N) (see Figure \ref{fig:gold-standard-generation}) for labeling the matches as duplicates or non-duplicates. The generated gold standard contains three duplicates\footnote{We do not label restaurant instances that have only one property value to compare, for example, only name.}. 2) we added manually twenty duplicates into the dataset and they were labeled as duplicates in our gold standard dataset. In total there are 23 duplicates.

\subsection{Similarity Metrics}
\label{subsec:similarity-metrics}
In this section, we describe different similarity metrics that are used by most of the tools described in section \ref{subsec:duplication-detection}. We group them as follows:

\begin{itemize}
    \item \textbf{String‐based metrics:} They are calculated by set‐based or edit‐distance‐based measurements to measure how similar two strings are. 
    \item \textbf{Vector Space metrics:} They are calculated based on the distance of two vectors in an ndimensional space. 
    \item \textbf{Point‐set distance metrics:} These are calculated by measuring the distance between two sets of points. 
    \item \textbf{Temporal similarity metrics:} These metrics are used to find the temporal relationship between two events. 
    \item \textbf{Topological similarity metrics:} These metrics are used to find out if two shapes on a plane are somehow related.
\end{itemize}

\subsection{Tool Selection}
\label{subsec:tool-selection}
In this section, we define criteria to selec the tools to be evaluated. We identified aroun 12 tools (see Section \ref{sec:literature}), which are detailed described in~\cite{FenselSAHKPTUW2020}. To reduce the number of tools to be evaluated, we introduced three criteria:
\begin{enumerate}
    \item The tool must be available online.
    \item The tool must be able to detect duplicates in one single dataset.
    \item The tool should handle multiple values of a property.
\end{enumerate}

Considering the criteria, the selected tool w.r.t. the issues defined in Section \ref{subsec:duplication-detection} is Duke for addressing issue 1, and FAGI and Sieve for tackling issue 2. In addition, we also evaluate the tools DuDe and Limes regarding their features and how they tackle DD.

\subsection{Tool Setup and Execution}
\label{subsec:tool-setup}

The process of running the selected tools consist of the following steps:
\begin{itemize}
    \item For the case of DuDe, Duke, and LIMES: 1) specify the input dataset in the config file, 2) define a setup for detecting duplicates (e.g. properties to be compared, similarity metrics, and threshold of acceptance), and 3) run the tool for generating an output dataset that contains the detected duplicates.
    \item For the case of FAGI and Sieve: 1) define the input dataset in the config file, 2) define a setup for fusing data (e.g. filter to be applied, properties to be compared, and fusion functions to resolve conflicting property values), and 3) run the tool for generating an output dataset that contains the instances with resolved property values and a threshold of confidence.    
\end{itemize}

The experiment has been executed on a Intel Core i7-8550U CPU 1.80Ghz (4 cores), 16GB of RAM, using Ubuntu 18.04.3 LTS 64-bit, Java 11.0.5, and Eclipse IDE 4.14.

\section{Results and Discussion}
\label{sec:results}
This section discusses the results obtained through this study, answers to each research question described in Section \ref{sec:evaluation}, proposing improvements for the development of duplication detection (DD) tools, and a DD workflow.

\subsection{Results}
\label{subsec:results}
We have executed the DD tools and the results are available under the following link \textcolor{blue}{https://bit.ly/DuplicationDetectionReport}, which describes the metrics and properties used to compare restaurant instances. Moreover, it describes their precision, recall, and F-measure. In the following, we sum up the results:

\begin{itemize}
    \item Across all used comparators, the best recall value \textit{0.3043} was obtained by comparing url or latitude and longitude properties. The best precision value \textit{0.6667} was obtained by using Cosine, DiceCoefficient, and Jaccard comparator to compare all properties.
    \item The total number of duplicates in our dataset is 23. However, the JaroWinkler comparator shows a large number of false positives.
    \item The average time of executing a task in DuDe takes \textit{2.167} seconds and \textit{1.857} seconds in Duke. Furthermore, executing the same task many times in DuDe and Duke has returned the same result.
    \item Reducing the threshold from \textit{0.9} to \textit{0.8} does not affect a lot the results.
\end{itemize}

In the following Section, we discuss more in detail the outcome of the tools by answering our research questions defined in Section \ref{sec:evaluation}.

\subsection{Answering the Research Questions}
\label{sec:answering-rq}

We tested the use of the tools w.r.t. their setup, execution, available preprocessing techniques, available measurement metrics, available features, and their performance. In the following, we answer our research questions.

\paragraph{RQ1: Are the identified tools able to detect duplicates in KGs?}
\label{sec:rq-1}
The DuDe tool configuration is based on creating a java main class where the input dataset, properties, comparators, and thresholds are defined. Dude was fed with a CSV file. So, we converted the input dataset described in Section \ref{subsec:dataset-selection} to CSV format. Moreover, DuDe has detected between 2 to 7 duplicates (it depends on the configuration) out of 23 duplicates (from our gold standard). DuDe can not handle null values, for example, cases where there is no value for a property.

The configuration file for running Duke and LIMES are written in XML format where the properties, comparators, threshold of acceptance, and datasets are declared. LIMES does not detected duplicates in one single dataset\footnote{LIMES is focused in record linkage across different datasets. However, its features can be applied to DD, e.g., the cleaners, comparators that it has implemented.}. Duke found 6 true positive duplicates by using DiceCoefficient comparator on all properties and 6 false positive duplicates. Also, Duke considers instances that have only one property to compare as duplicates, e.g., the \textit{schema:name} property\footnote{Instances that are described using only the \textit{schema:name} property have high chances to be selected as duplicates, since there are no more property values to compare with.}.

In summary, DuDe has limitation to process null values and is very format dependant, LIMES is focused on DD across two different datasets, and Duke has the best performance on detecting duplicates.

\paragraph{RQ2: Have the identified tools cleaning and/or preprocessing techniques?}
\label{sec:rq-2}
Preprocessing techniques aim to make comparison easier by normalizing the data, e.g., removing/replacing values that are not likely to indicate genuine differences like country code values: \footnotesize\texttt{+43} versus \footnotesize\texttt{0043}, or lower-/uppercase values.

DuDe has not preprocessing techniques enstead it generates statistics, e.g., the number of records contained in the dataset. Duke has a list of preprocessing functions grouped by: string cleaners; configurable cleaners; and parsing cleaners. LIMES has preprocessing functions categorised by simple (e.g. lowercase, number, replace) and complex (i.e. it allows concatenating multiple property values or splitting them). 

To sum up, a set of around 15 cleaner functions are found across the tools, which can help to prepare the dataset to achieve the DD. 

\paragraph{RQ3: Have the identified tools enough comparators?}
\label{sec:rq-3}
The identified tools have around 20 comparators (DiceCoefficient, JaccardIndex, JaroWinkler, Levenshtein, and more that can be used for comparing values. For instance, QGram comparator can be used on strings consisting of tokens that may be reordered (e.g. \footnotesize\texttt{HUGO’S BAR} and \footnotesize\texttt{BAR HUGO’S}). During execution time we observed that:
\begin{itemize}
    \item The property \textit{name} seems to be not representative for comparing duplicates. For instance, \footnotesize\texttt{HUGO'S TAPAS BAR} and \footnotesize\texttt{HUGO'S TRATTORIA \& PIZZERIA} are not the same restaurant, but they have similarity on using \footnotesize\texttt{HUGO'S}.
    
    \item The \textit{streetAddress} and \textit{url} property seem not to be discriminative, for instance, two restaurants (e.g. \footnotesize\texttt{HUGO's TAPAS BAR} and \footnotesize\texttt{HUGO'S TRATTORIA \& PIZZERIA}) have the same address, because there are placed in the same hotel.
    
    \item The values of \textit{url}, \textit{latitude}, and \textit{longitude} properties have shown to be representative (at least with the restaurant dataset) because they have the best recall (see Section \ref{sec:results}). Therefore, Geoposition comparators seem to be necessary. For instance, LIMES has a set of 10 geoposition comparators.
\end{itemize}

In summary, the tools have a similar number of comparators, and they are used for specific types of values. Note that, the results showed during the analysis only correspond to the dataset described in Section \ref{subsec:dataset-selection}.

\paragraph{RQ4: How the identified tools tackle conflicting property values?}
\label{sec:rq-4}

The tools focused on resolving conflicting property values are FAGI and Sieve. They consider some data quality metrics, such as timeliness and provenance of data to resolve the conflicting property values. The fusion strategy is defined through an XML configuration file, which defines the filters (e.g. domain restaurant), properties to be fused (e.g., name), and the fusion function to be applied. 

Some fusion functions described across those tools are: Filter that removes all values for which the input quality assessment is below a threshold; Average which takes the average value of all values for a given property; and Voting that takes the most frequent value for a given property.

To sum up, Sieve and FAGI have complementary fusion functions that can be used to resolve conflicting property values in KGs.

\paragraph{RQ5: How flexible is the configuration of the identified tools?}
\label{sec:rq-5}

The configuration file for the tools has to define the cleaners, comparators, input dataset, output file, properties, and thresholds of acceptance for properties. It needs to be written in a java file in case of DuDe and an XML file in case of Duke and LIMES. 

Duke has a feature called \textit{genetic algorithm} which given a dataset and a gold standard generates a proper configuration for comparing properties (i.e. passive mode). Furthermore, Duke can generate a gold standard by asking the user whether matches are correct or not (i.e. active mode).

Duke and LIMES can be used as APIs: allowing to process new and changed data (i.e. Duke); and implementing a RESTful API of configuration files and serving a browser frontend with a GUI to write the configuration file (i.e. LIMES).

\paragraph{RQ6: How scalable are the identified tools regarding large Knowledge Graphs?}
\label{sec:rq-6}

For measuring the scalability of the selected tools, we choose the dataset \texttt{1billion}\footnote{This dataset can be retrieved from \url{https://graphdb.sti2.at/repositories/1billion}} that contains 5,037,555 triples of restaurant instances. DuDe has been executed and it returned an \textit{OutOfMemoryError}, Duke has found 2,200 potential duplicates for a threshold greater than 0.9, and LIMES does not return any duplicate. Since LIMES is focus on linking two datasets, we take the dataset defined in Section \ref{subsec:dataset-selection} and the \texttt{1billion} dataset and it did not find any duplicate between them.

The size of a dataset matters. If we want to compare 10,000 records with each other, this process can be very slow because it would have a complexity of $O(n\textsuperscript{2})$. Duke uses a database interface where all records are indexed, and retrieves potential matches according to the index. Furthermore, Duke provides several different indexing implementation\footnote{See more: \url{https://github.com/larsga/Duke/wiki/DatabaseConfig}} to process datasets, for instance, LuceneDatabse (e.g. it indexes the records in a Lucene index), inMemoryDatabase (only applicable to small datasets), InMemmoryBlocking (using various KeyFunctions like name, address); and MapDB database\footnote{\url{http://www.mapdb.org/}}.

\subsection{Improvement Suggestions for Development of Duplication Detection (DD) Tools}
\label{sec:duplicate-guidelines}
This section provides improvement suggestions to develop DD tools. Based on running the tools selected in Section \ref{subsec:tool-selection}, we suggest the following improvements:

\begin{itemize}
    \item \textbf{Improving tool's usability:} the tool must minimize the required effort to be used and to understand its outputs, e.g., implement a GUI feature for exploring results and navigating between duplicates.
    \item \textbf{Showing status information:} the tool must show all log information about the task in hand like loading data, cleaning, etc.
    \item \textbf{Improving the selection of features for a dataset:} the tool must analyse the input dataset and report which properties perform better. e.g. removing redundant and irrelevant features can improve the detection quality.
    \item \textbf{Combining different metrics:} the tool must implement \footnotesize\texttt{AND}, \footnotesize\texttt{MIN}, \footnotesize\texttt{MAX}, and \footnotesize\texttt{OR} operators, which allow combining different metrics to improve the DD precision, e.g. \footnotesize\texttt{AND(trigrams(x.name,y.name), euclidean(x.lat|x.long, y.lat|y.long))}.
    \item \textbf{Providing results analysis:} the tool should provide statistics and graphical visualization about its performance, such as: accuracy; precision; recall; and F-measure. These statistics could be useful for comparison of different DD tools.
\end{itemize}

In short, we propose five improvements that can be implemented to the development of DD tools.

\subsection{Duplication Detection Workflow}
\label{sec:duplicate-workflow}
In this section, we describe the DD workflow as a semi-automatic process:
\begin{enumerate}
    \item \textbf{Selection of a Domain:} defines for which domain we want to do the DD. For instance, Hotel, Event, or Restaurant domain.
    \item \textbf{Build a gold standard:} allows a user labelling data as duplicates or non-duplicates. The purpose of building a gold standard is to train machine learning algorithms that recommend which features of a domain should be used to detect duplicates.
    \item \textbf{Feature-selection:} defines useful features for detecting same or different entities. A feature selection technique can remove redundant or irrelevant features., e.g. PostalCode is strong indicator of that two cities are different entities.
    \item \textbf{Data normalization:} preprocesses the input dataset in order to normalize the data w.r.t. lower/upper case, phoneNumber (e.g. \footnotesize\texttt{+43} == \footnotesize\texttt{0043}), and more.
    \item \textbf{Configure DD setup:} defines the configuration for detecting duplicates, for instance, a) manually, a user defines properties to compare, comparators, threshold of acceptance, or b) semi-automatically, the genetic algorithm (like shown by Duke) can generate an ideal configuration based on a gold standard.
    \item \textbf{Comparison method:} provides and defines comparison methods to find duplicates., e.g. Blocking methods, Sorted Neighborhood, and more. 
    \item \textbf{Run/train:} the DD tool runs over the input dataset and computes similarity values. Then, it trains a feature like \textit{genetic algorithm} that refines the DD setup.
    \item \textbf{Verify DD setup:} after running/training the tool its setup must be verified, i.e., the properties, comparators, and threshold of acceptance need to be adjusted.
    \item \textbf{Run the tool:} runs the tool over the corpus and computes similarity values. For example, executing the tools through a graphic user interface (GUI).
    \item \textbf{Duplicates entities viewer:} provides a GUI to verify found duplicates (e.g. for \textit{sim} $>$ 0.9) and confirm whether they are the same, different, or related.
    \item \textbf{Define fusion strategies:} defines fusion operators with the aim of creating unique representation of an entity, i.e. decides what to do based on similarity values.
    \item \textbf{Run Fusion:} runs the tool over the identified duplicates and resolves conflicting property values, e.g. evaluates if given values are the same.
    \item \textbf{Monitor fusion:} provides options to control the fusion process through a GUI.
\end{enumerate}

To sum up, this section described the results obtained from executing the DD tools and comparing their performance, e.g. Duke has shown a better performance. Besides, we found more than 15 preprocessing functions, 20 comparators, and several fusion functions across the evaluated tools. Moreover, we proposed improvement suggestions and a DD workflow that can lead to future DD tools implementations.

\section{Conclusion and Future Work}
\label{sec:conclusion}
We investigated DD methods and tools, selected ones, and compared their functionality and performance based on a common set of criteria. Through this study and to the aim of contributing to the DD community the following conclusions are achieved:

\begin{itemize}
    \item Throughout this study, it was possible to identify an efficient tool for DD that is Duke. However, it still needs to be improved since its last update was in 2017 and it was able to tackle only the task 1 (see Section \ref{subsec:duplication-detection}). Moreover, FAGI and Sieve can tackle the fusion of conflicting property values in KGs.
    \item Most of the DD tools (e.g. Duke, LIMES, Silk) offer various similarity metrics and the possibility of build complex similarity metrics for improving their performance. For instance, string similarity metrics seem to be the most used by DD tools, although, while such simple metrics can be used in many cases, complex metrics are necessary in complex DD cases.
    \item Aiming to contribute for the DD community, improvement suggestions and a workflow were proposed to support future implementations of DD tools. 
    \item On one hand, the tools provide sufficiency of tool documentation, on the other hand, we notice usability issues of tools (see Section \ref{sec:duplicate-guidelines}).
\end{itemize}

Although our paper has presented methods and tools, improvement suggestions, and workflow in DD, we believe that there is still work to do in this field. In the following, we point out our future work and open research questions.

Firstly, our next steps involve the development of a DD tool to tackle the detection of duplicates in the context of KGs and the implementation of the suggested features to improve its performance. 

Secondly, a more detailed analysis of fusion strategies will be performed, e.g. how to resolve conflicting property values effectively and efficiently.

Finally, we want to point out some future research directions; dynamic data, which formulates the temporal scoping problem between duplicates, e.g. entities' states change over time which may affect their property values; crowdsourced-based DD, which implicates a costly procedure to generate a gold standard, or to evaluate the results, e.g. people only verify matches with a high confidence score.

\bibliographystyle{splncs04}
\bibliography{main.bib}


\end{document}